\documentclass[aps,prd,preprint,superscriptaddress,tightenlines,nofootinbib]{revtex4}

\usepackage{graphicx}
\usepackage{dcolumn}
\usepackage{bm}

\begin{document}

\preprint{CLEO CONF 06-01}   

\title{Measurement of Upper Limits for $\Upsilon\to \gamma + {\cal R}$ Decays}

\thanks{Submitted to the 33$^{\rm rd}$ International Conference on High Energy
Physics, July 26 - August 2, 2006, Moscow}

\author{J.~L.~Rosner}
\affiliation{Enrico Fermi Institute, University of
Chicago, Chicago, Illinois 60637}
\author{N.~E.~Adam}
\author{J.~P.~Alexander}
\author{K.~Berkelman}
\author{D.~G.~Cassel}
\author{J.~E.~Duboscq}
\author{K.~M.~Ecklund}
\author{R.~Ehrlich}
\author{L.~Fields}
\author{R.~S.~Galik}
\author{L.~Gibbons}
\author{R.~Gray}
\author{S.~W.~Gray}
\author{D.~L.~Hartill}
\author{B.~K.~Heltsley}
\author{D.~Hertz}
\author{C.~D.~Jones}
\author{J.~Kandaswamy}
\author{D.~L.~Kreinick}
\author{V.~E.~Kuznetsov}
\author{H.~Mahlke-Kr\"uger}
\author{P.~U.~E.~Onyisi}
\author{J.~R.~Patterson}
\author{D.~Peterson}
\author{J.~Pivarski}
\author{D.~Riley}
\author{A.~Ryd}
\author{A.~J.~Sadoff}
\author{H.~Schwarthoff}
\author{X.~Shi}
\author{S.~Stroiney}
\author{W.~M.~Sun}
\author{T.~Wilksen}
\author{M.~Weinberger}
\affiliation{Cornell University, Ithaca, New York 14853}
% ...
\collaboration{CLEO Collaboration} 
\noaffiliation
\date{July 26, 2006}

\begin{abstract} 

Motivated by concerns regarding possible
two-body contributions to the recently-measured
inclusive $\Upsilon$(nS)$\to\gamma+X$
direct photon spectrum,
we report on a preliminary new study of exclusive radiative decays of the
$\Upsilon$ resonances into two-body final states ${\cal R}\gamma$, with
${\cal R}$ some resonant hadronic state
decaying into four or more charged particles.
Such two-body processes are not explicitly addressed in the
extant theoretical frameworks used to calculate the inclusive direct
photon spectrum.
Using data collected from the CLEO III detector at the Cornell Electron
Storage Ring, 
we present upper limits for such $\Upsilon({\rm 1S})$,
$\Upsilon({\rm 2S})$, and $\Upsilon({\rm 3S})$ two-body decays as a function
of the recoil mass $M_{\cal R}$.
Additionally,
we place upper limits on the cross-section for ${\cal R}$ production via
radiative return 
for center-of-mass energies 
just below the $\Upsilon({\rm 1S})$, $\Upsilon({\rm 2S})$, and $\Upsilon({\rm 3S})$ energies.    
The results presented in this document are preliminary.
\end{abstract}

\maketitle

\section*{Introduction}
Very recently, we extracted $\alpha_s$ from a 
measurement of the direct photon
spectra in $\Upsilon(1S,2S,3S)\rightarrow{gg\gamma}$ \cite{r:shawn}.  
To extrapolate beyond the experimentally accessible direct photon momentum 
region, those measurements relied on several
theoretical parameterizations of the expected photon momentum 
spectrum in the $\Upsilon$ system \cite{r:Field,r:SotoGarcia} to
obtain the total direct $\Upsilon\to gg\gamma$ decay width
relative to the dominant $\Upsilon\to ggg$ width. The ratio of
those widths can then be used to estimate the strong coupling constant
at the energy scale of the $\Upsilon$.
These theoretical parameterizations 
generally neglect 
the contribution to the photon momentum spectrum due to 
two-body decays, e.g. $\Upsilon\to gg\gamma\to {\cal R}\gamma$, with 
${\cal R}$ some resonant hadronic state.  

As noted in the $\alpha_s$ extraction, 
contributions from such possible two-body decays 
may result in a slight underestimate of the extracted value of $\alpha_s$.  
This systematic consideration in the $gg\gamma$ analysis motivated a 
search for $\Upsilon\to {\cal R}\gamma$.  We concern ourselves with 
high multiplicity ($\ge 4$ charged tracks) final states,
as we employ
the same hadronic event selection cuts in this analysis that we did
in the $gg\gamma$ analysis \cite{r:shawn}.
We note that, although two-body branching fractions have been observed
for, e.g., $\Upsilon$(1S)$\to\gamma f_2(1270)$ at the level of
$10^{-4}$, the fraction of $f_2(1270)$ decays into $\ge$4 charged
tracks is only $\approx 3\%$ \cite{r:pdg}. Our sensitivity
to the $f_2(1270)$ is additionally reduced in this analysis by
its large width (compared to our typical photon energy resolution).

The analysis, in general terms, proceeds as follows. After selecting a
high-quality sample of $e^+e^-$ annihilations into hadrons using the
hadronic event selection cuts of the previous analysis \cite{r:shawn}, we plot the
inclusive isolated photon spectrum in data taken at both
on-$\Upsilon$-resonance and off-$\Upsilon$-resonance energies (the latter
sample is used for systematic checks of the overall procedure). A two-body
radiative decay of the $\Upsilon$ will produce a monochromatic photon
in the lab frame; the momentum of the radiated photon is related to the
mass of the recoil hadron ${\cal R}$ via Equation~\ref{eq:conversion} ($x_\gamma=E_\gamma/E_{beam}$).

\begin{equation}
M_{\cal R}=2E_{beam}\sqrt{1-({x_\gamma^{\cal R}})^2}
\label{eq:conversion}
\end{equation}

Assuming that the intrinsic width of the recoil hadron
is much smaller than the experimental photon energy resolution, then the
measured radiative photon energy should be a Gaussian centered at the
momentum $x_\gamma^{\cal R}$. For a 1 GeV (4.5 GeV) recoil photon, this
implies a recoil mass typically much narrower than 20 MeV (60 MeV).
Not knowing \emph{a priori} the mass 
of the hadron
${\cal R}$, we therefore perform a set of fits to the $\Upsilon$(1S) photon
spectrum to a Gaussian signal, centered at the value $x_\gamma^{\cal R}$,
and with a resolution corresponding to the known CLEO-III electromagnetic
calorimeter resolution, atop smooth polynomial backgrounds, over the
$0.2<x_\gamma<1.0$ photon energy range.
\footnote{It should be noted that 
`bumps' in the inclusive photon spectrum can be due not only to
resonant two-body decays but also to 
continuum threshold effects like 
the crossing of the $c\overline{c}$ threshold.
In the previous analysis, we also identified an excess of photons in
data as $x_\gamma\to1$. Further examination of these events indicated that
they were dominated by $e^+e^-\to\gamma\pi^+\pi^-\pi^+\pi^-$, although
the possibility that the 4-pion state resulted from the decay
of an intermediate resonance ${\cal R}$ was not investigated.}

We construct 95\% confidence level upper limits from these fits by adding 
$1.645*\sigma_A(x_\gamma)$ to the $A(x_\gamma)$ distribution,
where $A(x_\gamma)$ is the $x_\gamma$-dependent Gaussian fit area
and $\sigma_A(x_\gamma)$ is the fit error.
We then recast this upper limit as a function of recoil mass $M_{\cal R}$,
correct for the efficiency loss due to the fiducial acceptance of 
the detector (for the purposes 
of this correction, it is assumed that ${\cal R}$ is pseudo-scalar with 
a corresponding $1+\cos^2\theta$ angular distribution for the
recoiling $\gamma$) and of the event 
and photon selection cuts that define our data sample.  Note that the
exact form of the efficiency correction due to the event and photon 
selection cuts varies with the decay final states considered for 
${\cal R}$.  To be conservative, we derive our 
$x_\gamma$-dependent efficiency correction 
from the decay mode yielding the lowest 
reconstruction efficiency imaginable.

\begin{figure}[htpb]
\centerline{\includegraphics[width=8cm]{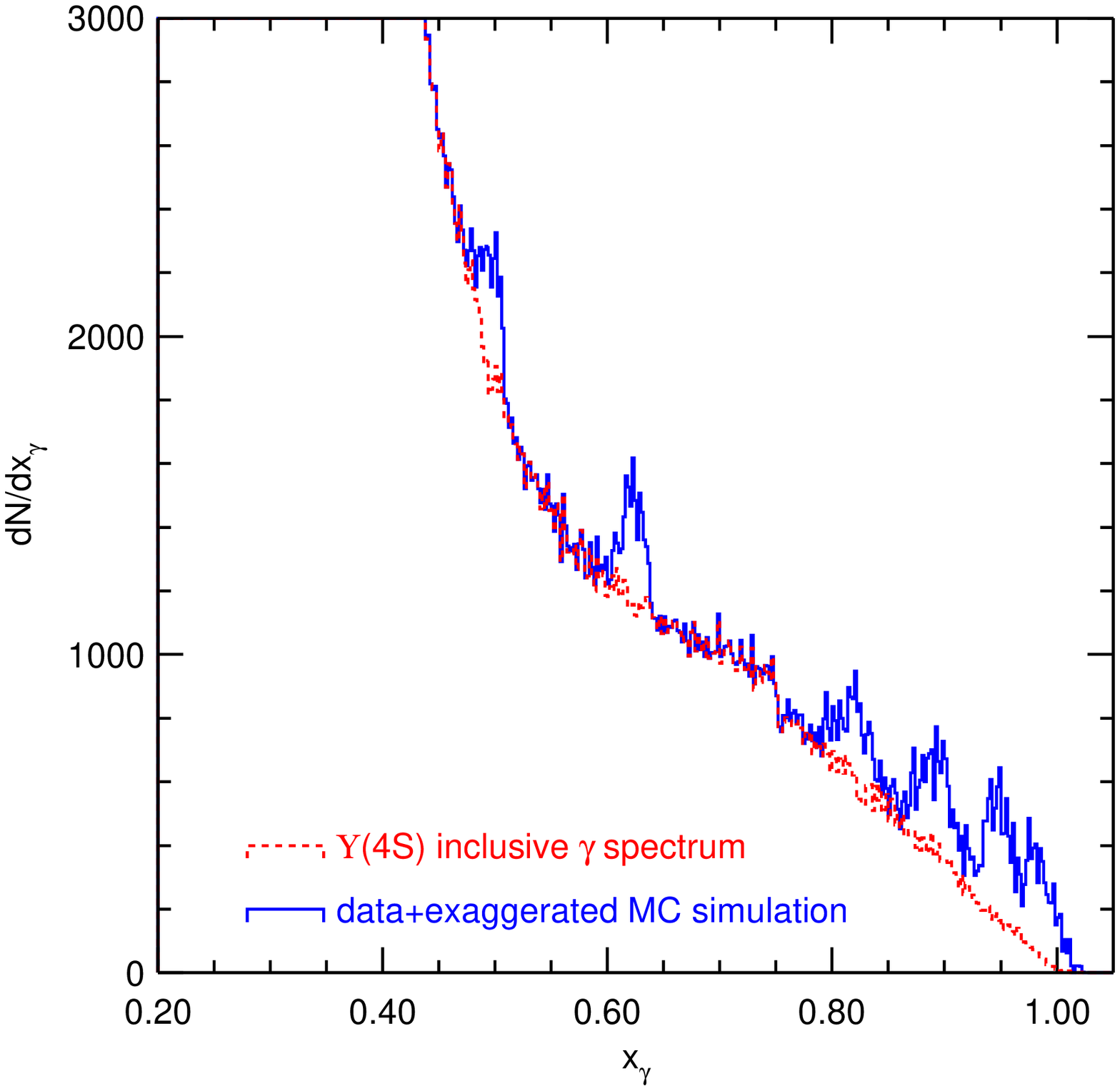}}
\caption{$\Upsilon(4S)\to {\cal R}\gamma$, ${\cal R}\to 4\pi$, for
various hypothetical ${\cal R}$ masses.  The magnitude of 
${\cal B}$($\Upsilon(4S)\to {\cal R}\gamma$, ${\cal R}\to 4\pi$) 
has been grossly exaggerated for the sake of presentation.
The red (dotted) curve is 
$\Upsilon(4S)$ data, while the blue (solid) curve is $\Upsilon(4S)$ data with signal
Monte Carlo added on top.  From the left, the first `bump' corresponds to a 
${\cal R}$ of mass $\approx7.6$ GeV, the second from the left corresponds to a 
${\cal R}$ of mass $\approx6.5$ GeV, the third from the left corresponds to a 
${\cal R}$ of mass $\approx4.7$ GeV and the fourth from the
left corresponds a ${\cal R}$ of mass $\approx3.3$ GeV.}
\label{fig:fakeSignal}
\end{figure}

This final efficiency-corrected limit is converted into an
$M_{\cal R}$-dependent branching ratio upper limit 
${\cal B}(\gamma {\cal R})$ by dividing 
the resulting yield by the
calculated total number of resonant $\Upsilon$ events.  For the off-resonance 
running, the distributions are divided by the off-resonance luminosity 
for comparison's sake.
An example of simulated signal
superimposed on background is given in 
Figure~\ref{fig:fakeSignal} for $\Upsilon(4S)\to\gamma+{\cal R}$, 
${\cal R}\to\pi^+\pi^-\pi^+\pi^-$.

\section*{Event Selection}
Event selection criteria in this analysis are identical to those imposed
in the previous analysis \cite{r:shawn}. The inclusive photon
spectra are therefore identical to those taken from our previous
analysis, as well. The background shape is approximately
exponential in the region from $0.2<x_\gamma<1.0$.

\section*{Fitting the Inclusive Photon Spectrum}

To extract the possible magnitude of a two-body
radiative signal, we step along the inclusive photon 
spectrum over the interval $0.2<x_\gamma<1.0$, fitting it to a Gaussian with width
equal to the detector resolution at that value of photon
energy, plus a background
parametrized by a smooth Chebyshyev polynomial.  
We assume that the intrinsic width of the resonance ${\cal R}$ is
considerably smaller than the detector resolution.
Our step size is
determined by the energy resolution of the detector $\sigma_{\rm E}$; we use steps of width 
$\sigma_{\rm E}/2$.
For photons in the central
``barrel'' region of the CsI 
electromagnetic calorimeter, at energies
greater than 2 GeV, the energy resolution
is given by
\begin{equation}
\frac{ \sigma_{\rm E}}{E}(\%) = \frac{0.6}{E^{0.73}} + 1.14 - 0.01E,
                                \label{eq:resolution1}
\end{equation}
where $E$ is the shower energy in GeV. At 100 MeV, the
calorimetric performance is about 20\% poorer than indicated by this expression
due to the material in front of the calorimeter itself.  

At each step, we use a
$\pm$10$\sigma_{\rm E}$ fitting window; the background
is expected to be relatively smooth over such a limited interval.  
We use a 3rd-(2nd-)order Chebyschev polynomial in the photon interval
$x_\gamma<(>)0.6$.
For each fit, the Gaussian fit area $A(x_\gamma)$ and fit error $\sigma_A(x_\gamma)$
is recorded. 
Note that the fits are highly correlated point-to-point, and
that the bin width is much finer than the detector resolution. 
Note also that
we have not attempted to analytically 
correct for smearing-induced distortions that may
result from the finite resolution of the detector; we simply
assume that the smeared spectrum can be well-described by a smooth higher-order
polynomial.

\section*{Extracting Upper Limits}

\begin{figure}[htpb]
\centerline{\includegraphics[width=8cm]{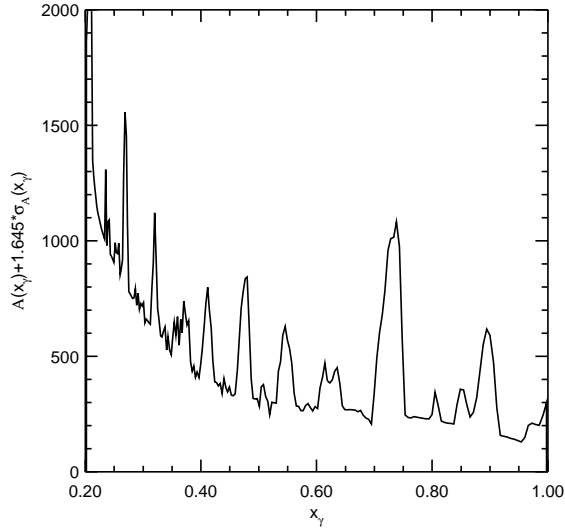}}
\caption{$A(x_\gamma)+1.645*\sigma_A(x_\gamma)$ versus $x_\gamma$ for fits to the 
$\Upsilon(1S)$ inclusive photon spectrum, where negative points have been mapped to 
$1.645*\sigma_A(x_\gamma)$.}
\label{fig:rawUpperLimit1S}
\end{figure}

To convert the $A(x_\gamma)$ distribution obtained from fitting the inclusive photon spectrum
into a one-sided upper 95\% confidence interval
upper limit, we add $1.645*\sigma_A(x_\gamma)$ point-wise to the 
$A(x_\gamma)$ distribution,
as a function of photon momentum.  In this process, since we are interested in
enhancements in the inclusive photon spectrum, all negative areas from the raw fits are
set equal to $1.645*\sigma_A(x_\gamma)$
at these points.  The resulting contour for the $\Upsilon(1S)$ fitting is shown in
Figure~\ref{fig:rawUpperLimit1S}.

We convert the limits as a function of 
photon energy $x_\gamma$
into a function of a hypothetical resonance recoil mass $M_{\cal R}$ via
Equation~\ref{eq:conversion}.  For the purposes of this conversion the mean values
for each running period of $E_{beam}$ 
are used for each data sample; we neglect the MeV-scale variation in
beam energies for a particular run period.  These values are given in Table~I.

\begin{table}[htpb]
\label{tab:beamEnergies}
\begin{center}
\begin{tabular}{|c|c|}\hline
Event Type & Average Beam Energy ($\overline{E_{beam}}$)  \\
\hline
Resonance $\Upsilon(1S)$ & 4.73 GeV \\
Resonance $\Upsilon(2S)$ & 5.01 GeV \\
Resonance $\Upsilon(3S)$ & 5.18 GeV \\
Continuum $\Upsilon(1S)$ & 4.72 GeV \\
Continuum $\Upsilon(2S)$ & 5.00 GeV \\
Continuum $\Upsilon(3S)$ & 5.16 GeV \\
Continuum $\Upsilon(4S)$ & 5.27 GeV \\
\hline
\end{tabular}
\end{center}
\caption{The mean values of $E_{beam}$ for each data sample used in this analysis,
used to map our upper limit contour from a function of $x_\gamma$ to a function of 
$M_{\cal R}$.}
\end{table}

\begin{figure}[htpb]
\centerline{\includegraphics[width=8cm]{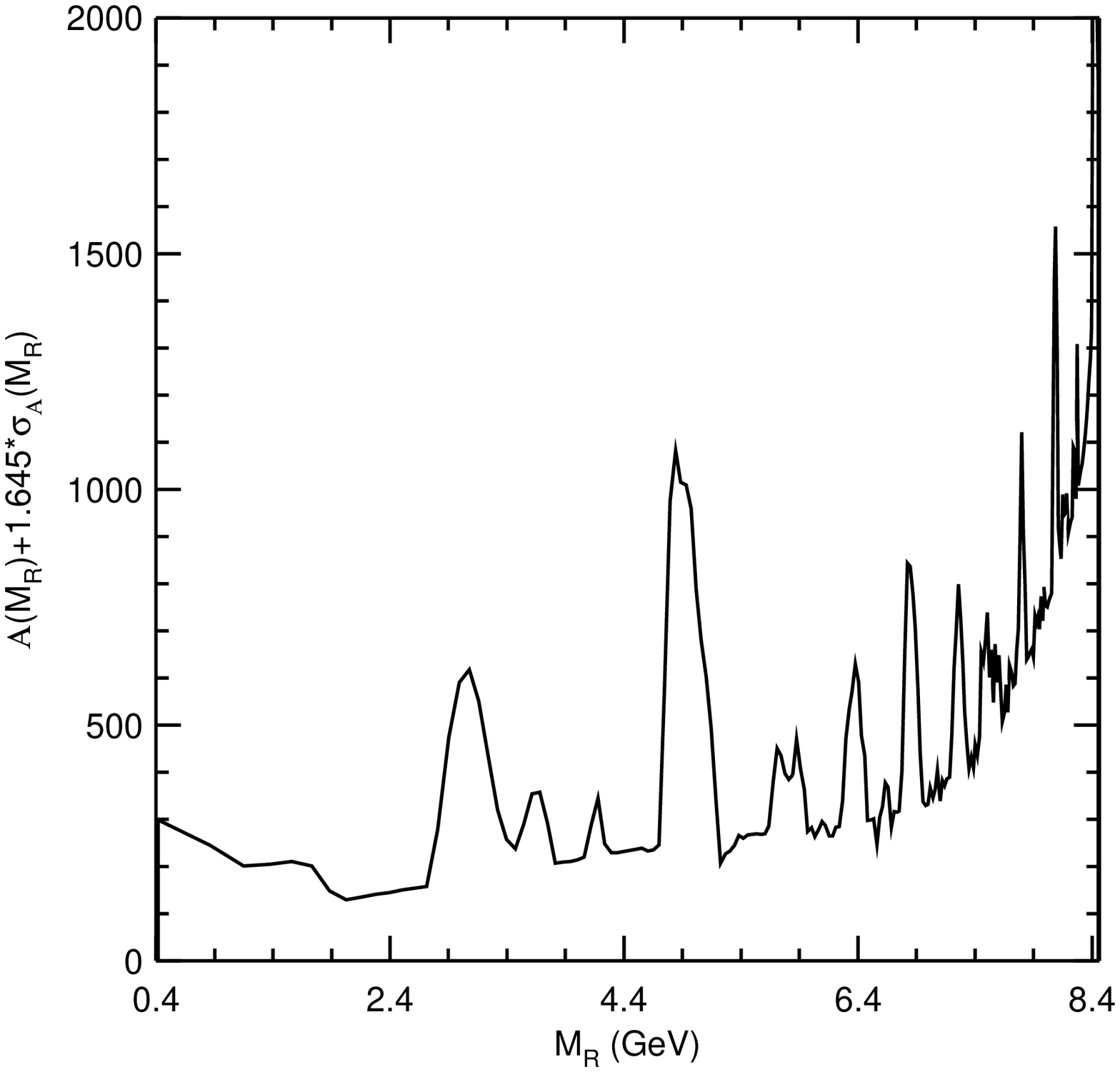}}
\caption{$A(M_{\cal R})+1.645*\sigma_A(M_{\cal R})$ versus $M_{\cal R}$ for fits to the 
$\Upsilon(1S)$ inclusive photon spectrum.  This curve was obtained from mapping the points in 
Figure~\ref{fig:rawUpperLimit1S} into recoil mass space via Equation~\ref{eq:conversion}}  
\label{fig:rawRecoilUpperLimit1S}
\end{figure}

The resulting $M_{\cal R}$-dependent contour, 
for the $\Upsilon(1S)$, is shown in Figure~\ref{fig:rawRecoilUpperLimit1S}.

\section*{Efficiency Correction}

We consider two efficiency corrections to the upper limit contour:
one due to the fiducial acceptance of the 
detector, and the other due to our event and shower selection cuts.

\begin{figure}[htpb]
\centerline{\includegraphics[width=8cm]{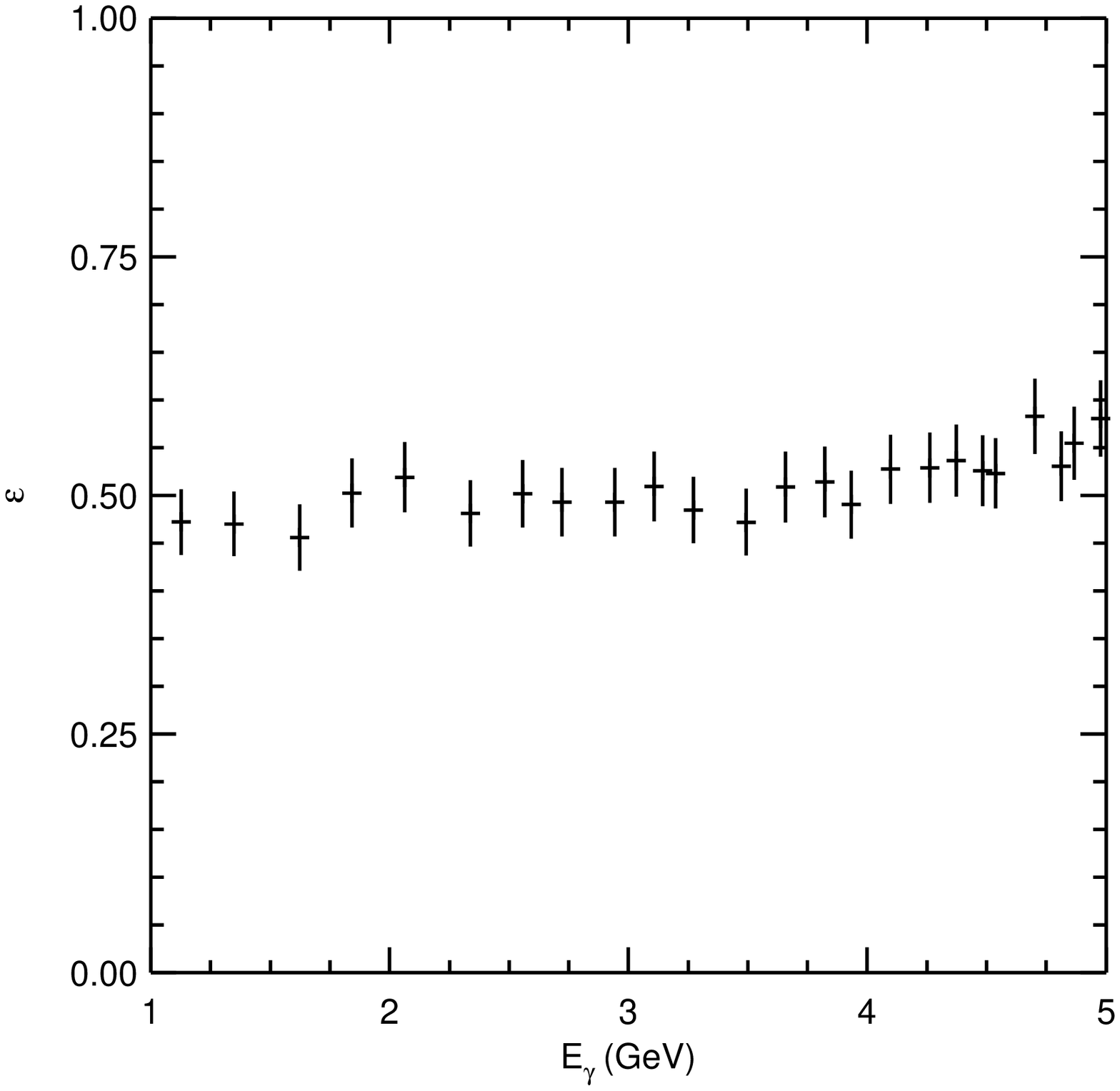}}
\caption{The efficiency for detecting an $\Upsilon\to\gamma+{\cal R}$, 
${\cal R}\to4K\pi^0$ event as a function of observed photon energy $E_\gamma$.
Each point in this efficiency is obtained from a different 
${\cal R}$ mass hypothesis.  This photon momentum-dependent efficiency 
correction distribution is used to point-wise correct our upper limit, where the efficiency between
points in this distribution is estimated by linear interpolation.}
\label{fig:4Kpi0Efficiency}
\end{figure}

For photons in the ``barrel'' of the detector ($|\cos\theta|<0.707$, with
$\theta$ the polar angle of the photon momentum relative to the beam axis), we assume that 
${\cal R}$ is a pseudo-scalar, which corresponds to a $1+\cos^2\theta$ distribution on the 
photons in the two-body decays we are considering. This choice is 
arbitrary and amounts to an $\approx0.6$
uniform acceptance efficiency correction to our limit.  In addition to this 
angular acceptance correction,
we assess an efficiency correction due to the CLEO~III detection efficiency.
Not knowing \emph{a priori} what the decay mode of our hypothetical 
resonance ${\cal R}$ will be,
we have
generated 5000-event Monte Carlo samples spanning a wide range of final
state charged (and neutral) multiplicities
and masses $M_{\cal R}$.
In the interests of producing a conservative upper limit, 
we take our efficiency correction from the mode which yields the lowest average efficiency 
in the main $M_{\cal R}$ interval of interest.  
A list of ${\cal R}$ decay modes considered in this analysis and their average 
efficiencies (averaged over the photon momentum spectrum from 
$1.0$ GeV $\leq{E_\gamma}\leq4.5$ GeV) is given in Table~II.  
We thereby derive a collection of efficiency measurements distributed over
photon momentum $E_\gamma$, each of which corresponds to a different hypothesis for the mass 
$M_{\cal R}$.  To obtain efficiencies between these points (where we have not explicitly filled
in the efficiency with a $M_{\cal R}$ hypothesis), we perform a linear extrapolation between
the two neighboring points.  In this manner, we point-wise efficiency correct our upper limit 
as a function of $x_\gamma$ (or $E_\gamma$) 
before mapping the upper limit into $M_{\cal R}$.
We find that the lowest efficiency for various 
hypothetical decay modes of ${\cal R}$ considered was obtained from ${\cal R}\to4K\pi^0$
(Figure~\ref{fig:4Kpi0Efficiency}).  We therefore use this efficiency 
function to determine our upper limit contours.

\begin{table}[htpb]
\label{tab:eventeff}
\begin{center}
\begin{tabular}{|c|c|}\hline
Event Type & Average Efficiency ($\overline{\epsilon}$)  \\
\hline
${\cal R}\to{K^+K^-\pi^+\pi^-}$ & 0.53 $\pm$ 0.03 \\
${\cal R}\to{K^+K^-\pi^+\pi^-\pi^0}$ & 0.53 $\pm$ 0.02 \\
${\cal R}\to{K^+K^-\pi^+\pi^-\pi^0\pi^0}$ & 0.54 $\pm$ 0.02 \\

${\cal R}\to{K^+K^-p^+p^-}$ & 0.56 $\pm$ 0.02 \\
${\cal R}\to{K^+K^-p^+p^-\pi^0}$ & 0.50 $\pm$ 0.05 \\
${\cal R}\to{K^+K^-p^+p^-\pi^0\pi^0}$ & 0.57 $\pm$ 0.02 \\

${\cal R}\to{p^+p^-\pi^+\pi^-}$ & 0.62 $\pm$ 0.03 \\
${\cal R}\to{p^+p^-\pi^+\pi^-\pi^0}$ & 0.54 $\pm$ 0.05\\
${\cal R}\to{p^+p^-\pi^+\pi^-\pi^0\pi^0}$ & 0.63 $\pm$  0.02 \\

${\cal R}\to{K^+K^-K^+K^-}$ & 0.50 $\pm$ 0.02 \\
${\cal R}\to{K^+K^-K^+K^-\pi^0\pi^0}$ & 0.49 $\pm$  0.02 \\

${\cal R}\to{p^+p^-p^+p^-}$ & 0.67 $\pm$ 0.02 \\
${\cal R}\to{p^+p^-p^+p^-\pi^0}$ & 0.65 $\pm$  0.02 \\
${\cal R}\to{p^+p^-p^+p^-\pi^0\pi^0}$ & 0.63 $\pm$  0.02 \\

${\cal R}\to{\pi^+\pi^-\pi^+\pi^-}$ & 0.59 $\pm$ 0.02 \\
${\cal R}\to{\pi^+\pi^-\pi^+\pi^-\pi^0}$ & 0.65 $\pm$ 0.02 \\
${\cal R}\to{\pi^+\pi^-\pi^+\pi^-\pi^0\pi^0}$ & 0.59 $\pm$ 0.01 \\

${\cal R}\to{\pi^+\pi^-\pi^+\pi^-4\pi^0}$ & 0.57 $\pm$ 0.02 \\
${\cal R}\to{\pi^+\pi^-\pi^+\pi^-6\pi^0}$ & 0.60 $\pm$ 0.02 \\
${\cal R}\to{\pi^+\pi^-\pi^+\pi^-8\pi^0}$ & 0.60 $\pm$ 0.02 \\

${\cal R}\to{K^+K^-K^+K^-K^+K^-}$ & 0.68 $\pm$ 0.04 \\
${\cal R}\to{p^+p^-p^+p^-p^+p^-}$ & 0.53 $\pm$ 0.04 \\
${\cal R}\to{\pi^+\pi^-\pi^+\pi^-\pi^+\pi^-}$ & 0.74 $\pm$ 0.03 \\ \hline
${\cal R}\to{K^+K^-K^+K^-\pi^0}$ & {\bf 0.48 $\pm$  0.02} \\ \hline
\end{tabular}
\end{center}
\caption{Average efficiencies for the reconstruction of 
various events that could be detected in this analysis, obtained by
fitting the photon momentum-dependent reconstruction efficiencies to a straight line
in the interval $1.0$ GeV $<E_\gamma<$ 4.5 GeV.  The presented errors 
on the efficiencies are statistical only. The lowest-efficiency
final state ($K^+K^-K^+K^-\pi^0$) is used for setting upper limits.}
\end{table}

\section*{Results}

\begin{figure}[htpb]
\centerline{\includegraphics[width=8cm]{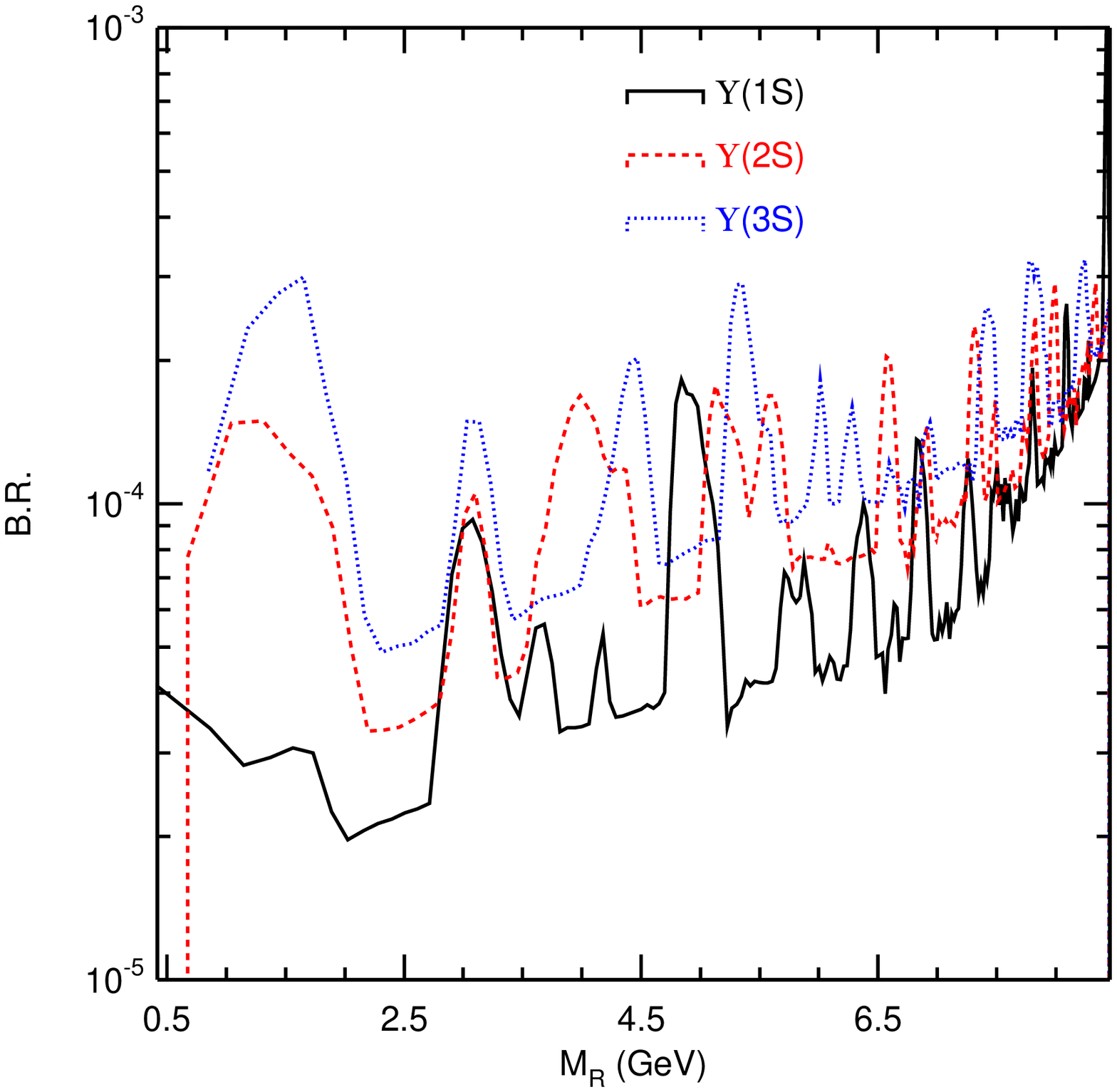}}
\caption{The $M_{\cal R}$-dependent ${\cal B}(\gamma {\cal R})$ upper limit contours obtained for 
$\Upsilon\to\gamma+{\cal R}$, ${\cal R}\to\geq4$ charged tracks for the 
$\Upsilon(1S)$, $\Upsilon(2S)$ and $\Upsilon(3S)$.  All limits are of order ${\cal B}(\gamma {\cal R})\approx10^{-4}$.}
\label{fig:resonanceLimit}
\end{figure}

\begin{figure}[htpb]
\centerline{\includegraphics[width=8cm]{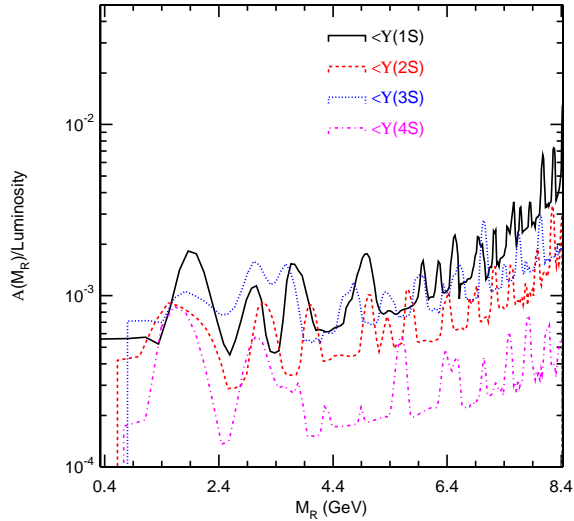}}
\caption{Our $M_{\cal R}$-dependent cross-section upper limit contours obtained for 
$e^+e^-\to\gamma+{\cal R}$, ${\cal R}\to\geq4$ charged tracks for the continua below 
$\Upsilon(1S)$, $\Upsilon(2S)$, $\Upsilon(3S)$ and $\Upsilon(4S)$ 
(nb).  
This plot 
is obtained by dividing the result of our fitting procedure 
on the continuum by the off-resonance
luminosity. The angular correction here is (as before) 
$1+\cos^2\theta_\gamma$.}
\label{fig:offResonanceLimit}
\end{figure}

To convert the efficiency corrected upper limit contour into an
upper limit on the two-body radiative branching ratio ${\cal B}(\gamma {\cal R})$,
we simply divide the efficiency-corrected 
upper limit contour by the total
calculated number of $\Upsilon(1S)$, $\Upsilon(2S)$ and $\Upsilon(3S)$ events.  These numbers
were calculated in our previous analysis \cite{r:shawn}, and for convenience are 
reproduced in Table~III.  For comparison's sake, the efficiency-corrected upper 
limit contours obtained from fitting below $\Upsilon$ resonance data have been divided by
the total luminosity of off-resonance data taking and therefore correspond 
to an upper limit on a cross-section. 
The resulting on-resonance 
upper limit ${\cal B}(\gamma {\cal R})$ is shown in Figure~\ref{fig:resonanceLimit}, and
the resulting off-resonance cross-section
``limit'' is shown in Figure~\ref{fig:offResonanceLimit}.
To set the scale of the continuum cross-section limits,
the raw ISR cross-section for $e^+e^-\to\psi+\gamma$ is 
expected to be $\sim$5 pb in the 10 GeV center-of-mass
regime, 
implying an expected observed cross-section into $\ge$4 charged
tracks $\sim 10^{-4}$ nb, taking into account the
strong forward peaking expected for ISR processes.

\begin{figure}[htpb]
\centerline{\includegraphics[width=8cm]{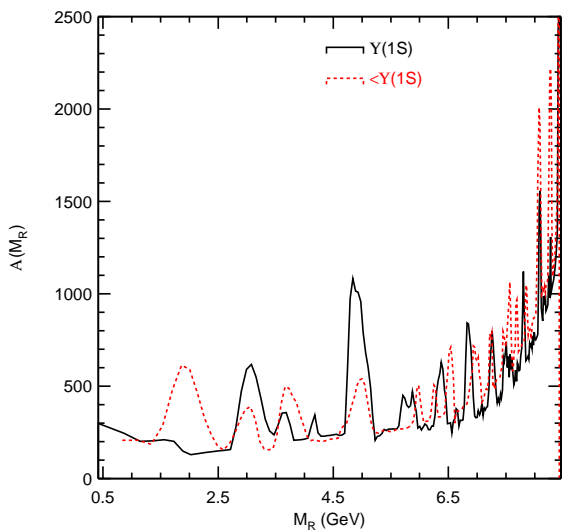}}
\caption{Comparison of the $M_{\cal R}$-dependent Gaussian fit area $A(M_{\cal R})$
for the $\Upsilon(1S)$ versus the continuum below $\Upsilon(1S)$.  Note the correlation
between the resonance and the below-resonance structure.  Note also that the below-resonance upper limit curve has been scaled to match the 'floor' of the on-resonant upper limit curve.}
\label{fig:limit1sVb1s}
\end{figure}

\begin{figure}[htpb]
\centerline{\includegraphics[width=8cm]{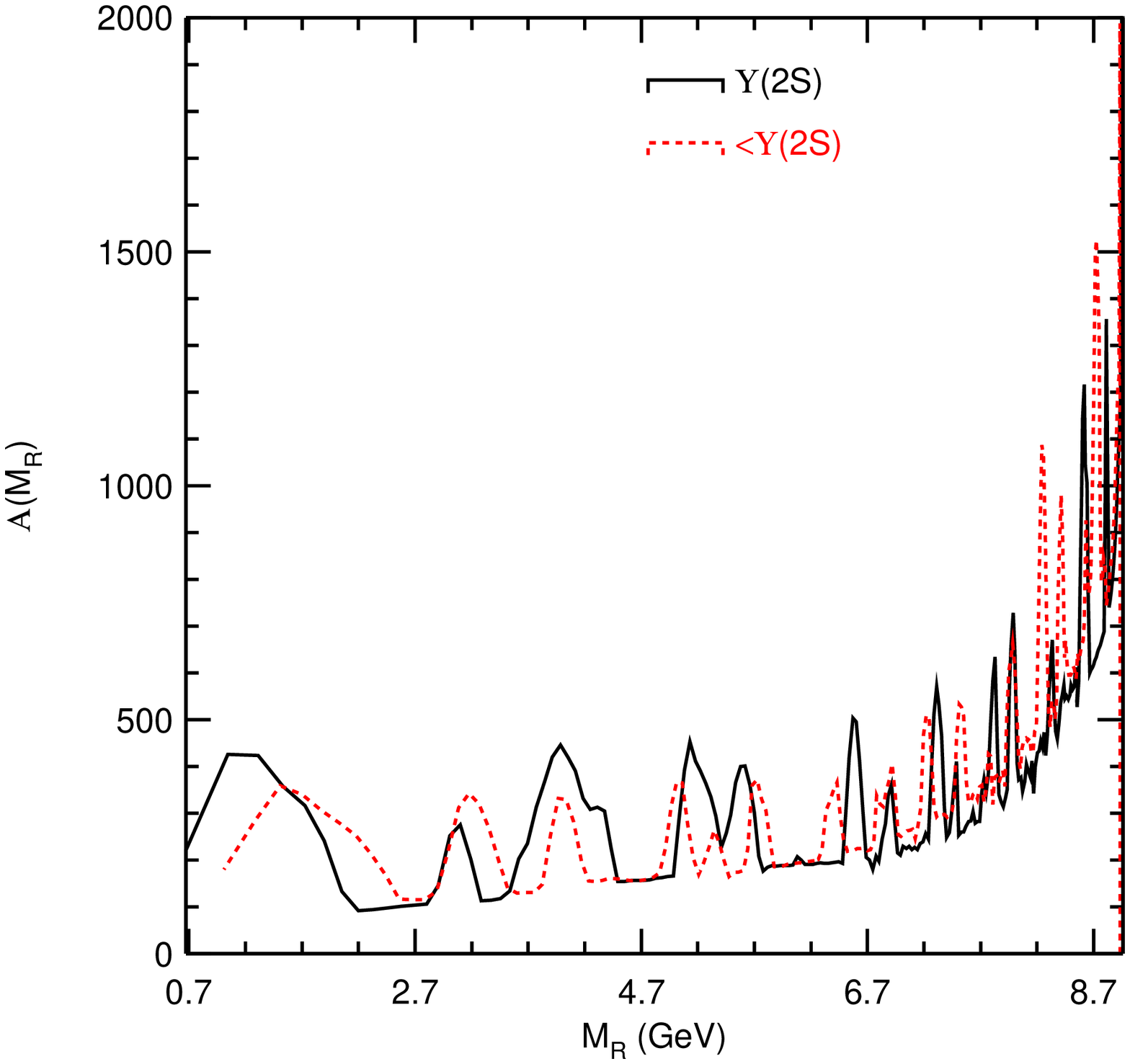}}
\caption{Comparison of the $M_{\cal R}$-dependent Gaussian fit area $A(M_{\cal R})$
for the $\Upsilon(2S)$ versus the continuum below $\Upsilon(2S)$.  Note that the below-resonance upper limit curve has been scaled to match the 'floor' of the on-resonant upper limit curve.}
\label{fig:limit2sVb2s}
\end{figure}

\begin{figure}[htpb]
\centerline{\includegraphics[width=8cm]{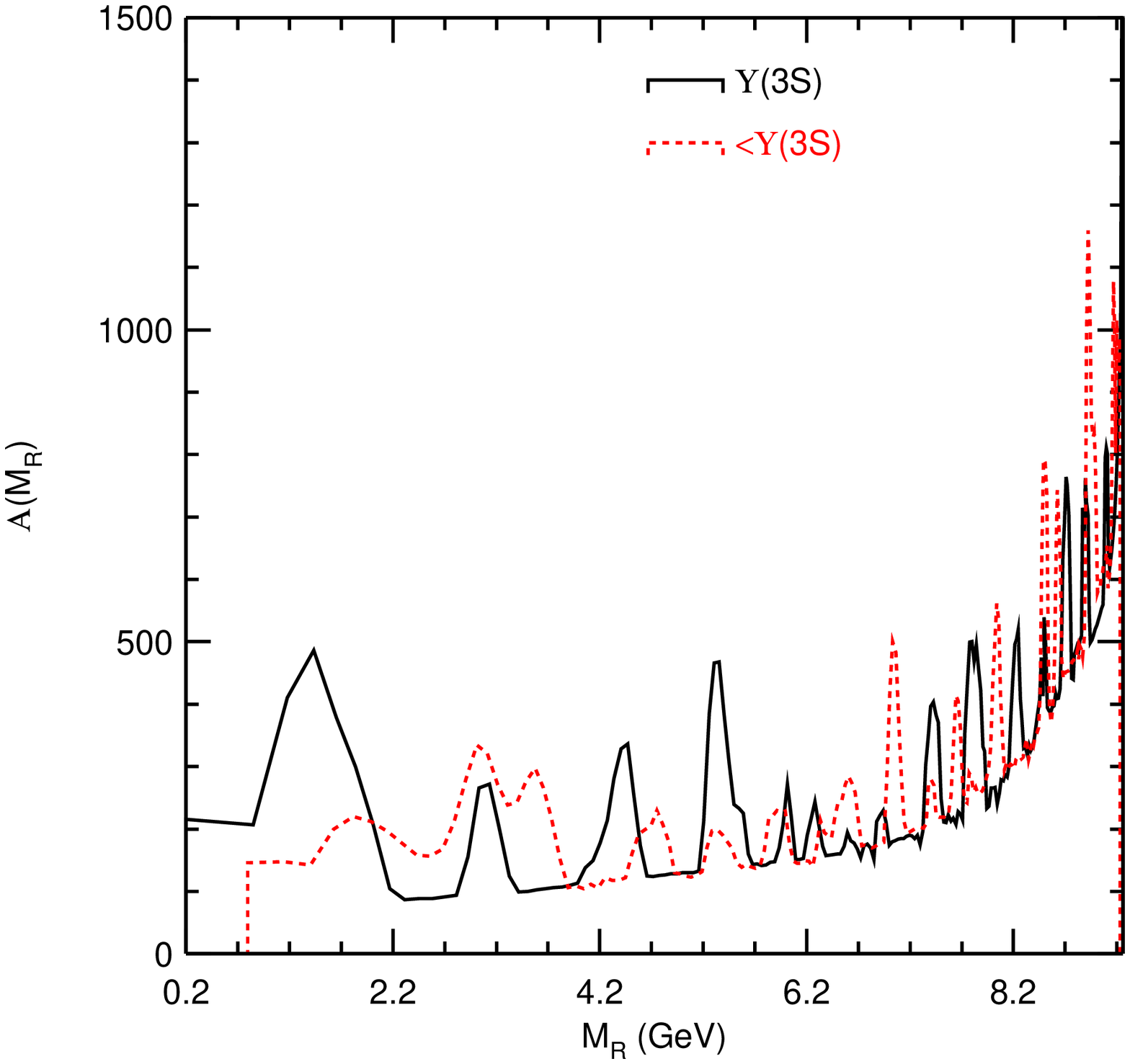}}
\caption{Comparison of the $M_{\cal R}$-dependent Gaussian fit area $A(M_{\cal R})$
for the $\Upsilon(3S)$ versus the continuum below $\Upsilon(3S)$.  Note that the below-resonance upper limit curve has been scaled to match the 'floor' of the on-resonant upper limit curve.}
\label{fig:limit3sVb3s}
\end{figure}

Given the fact that we have not performed a continuum 
subtraction on the on-resonance inclusive photon
spectrum from $\Upsilon$ decays, it is interesting to compare the structure
observed in Figure~\ref{fig:resonanceLimit} with structure observed when we
apply the fitting procedure to continuum data.
Figures~\ref{fig:limit1sVb1s}, \ref{fig:limit2sVb2s} and \ref{fig:limit3sVb3s} 
show the resonances limits of Figure~\ref{fig:resonanceLimit} separately, with the 
respective continuum limits of Figure~\ref{fig:offResonanceLimit} overlaid for comparison.
We observe some, albeit not complete, correlation between the two. 

\begin{table}[htpb]
\label{tab:eventNumber}
\begin{center}
\begin{tabular}{|c|c|}\hline
$\Upsilon$ Resonance & $N_{\rm total}(\Upsilon$(nS)) $(\times 10^6)$ \\
\hline
$\Upsilon$(1S) & 21.0 $\pm$ 0.06\\
$\Upsilon$(2S) & 8.4 $\pm$ 0.04\\
$\Upsilon$(3S) & 5.2 $\pm$ 0.06\\
\hline
\end{tabular}
\end{center}
\caption{The total number of calculated $\Upsilon$(1S), 
$\Upsilon$(2S) and $\Upsilon$(3S) events in our data samples \cite{r:shawn}.}  
\end{table}

\section*{Cross-Check}

\begin{figure}[htpb]
\centerline{\includegraphics[width=8cm]{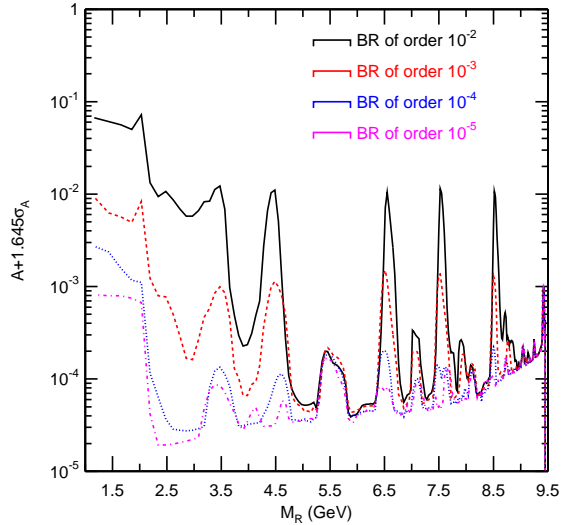}}
\caption{Upper limit contours derived from applying our procedure to 
fabricated Monte Carlo signal spectra.  We reconstruct all input signals above our upper limit floor ($\approx10^{-4}$) that are within our accessible recoil mass range.}
\label{fig:fake_signal_upper_limit}
\end{figure}

In order to ensure that we are able to identify a 
signal at a given sensitivity level, we embedded pure Monte Carlo signal in data, and 
performed our fitting procedure on the resulting 
distribution in order to ensure that we recovered the correct signal magnitude in 
our branching ratio upper limit.  To do this, $\Upsilon(4S)\to\gamma+{\cal R}$, 
${\cal R}\to\pi^+\pi^-\pi^+\pi^-$ events were embedded in the $\Upsilon(4S)$ inclusive
photon spectrum with branching ratios of the order of $10^{-5}$, $10^{-4}$, $10^{-3}$, 
and $10^{-2}$, under $10$ different $M_{\cal R}$ hypotheses:
$M_{\cal R}=0.6$ GeV, $1.5$ GeV, $2.5$ GeV, $3.5$ GeV, $4.5$ GeV, $6.5$ GeV, 
$7.5$ GeV, $8.5$ GeV, $9.5$ GeV and $10.5$ GeV.
The resulting upper limit contours
derived from applying our procedure to these spectra are show in 
Figure~\ref{fig:fake_signal_upper_limit}.
We reconstruct all signals above our upper limit floor (around $10^{-4}$) that
are within our accessible recoil mass range.

\section*{Systematic errors}
We identify and account for systematic errors as follows:

\begin{enumerate}
\item We account for possible
systematics in our event and shower reconstruction efficiency by using 
the lowest-efficiency final state considered.
\item We assess a uniform 1\% degradation of the limit due to the luminosity uncertainty
as calculated in the previous analysis \cite{r:shawn}.
\item We degrade the limit uniformly by the uncertainty in the calculated number of total $\Upsilon(1S)$,
$\Upsilon(2S)$ and $\Upsilon(3S)$ events.  
\end{enumerate}

\subsection*{Summary}
As shown in Figure~\ref{fig:resonanceLimit}, our sensitivity
is of order $10^{-4}$ across all possible values of $M_{\cal R}$,
well above the threshold for any known
branching ratio for $\Upsilon\to\gamma$+pseudoscalar, pseudoscalar$\to{h^+h^-h^+h^-}$+neutrals.  
We measure upper limits of 
${\cal B}(\Upsilon(1S)\to\gamma+{\cal R}, {\cal R}\to\geq4$ charged tracks $) < 1.05\times10^{-3}$,
${\cal B}(\Upsilon(2S)\to\gamma+{\cal R}, {\cal R}\to\geq4$ charged tracks $) < 1.65\times10^{-3}$ and 
${\cal B}(\Upsilon(3S)\to\gamma+{\cal R}, {\cal R}\to\geq4$ charged tracks $) < 2.47\times10^{-3}$ for
all possible masses $M_{\cal R}$, under the assumption that ${\cal R}$ is a pseudoscalar.
Constraining $1.5$ GeV $<M_{\cal R}<5.0$ GeV we set a much more stringent limit of 
${\cal B}(\Upsilon(1S)\to\gamma+{\cal R}, {\cal R}\to\geq4$ charged tracks $) < 1.82\times10^{-4}$,
${\cal B}(\Upsilon(2S)\to\gamma+{\cal R}, {\cal R}\to\geq4$ charged tracks $) < 1.69\times10^{-4}$ and 
${\cal B}(\Upsilon(3S)\to\gamma+{\cal R}, {\cal R}\to\geq4$ charged tracks $) < 3.00\times10^{-4}$.
Additionally, we report these upper limits as a function of the
mass recoiling against the photon, as shown
in Figure~\ref{fig:resonanceLimit}.  We also report cross-section 
limits for resonant processes
on the continuum (Figure~\ref{fig:offResonanceLimit}).

We note that we appropriately ignored the 
distortion of the inclusive photon spectrum in our previous extraction of $\alpha_s$ due to 
two-body decays as we limit the branching ratio of these events to be $\le 10^{-4}$.
Further work on exclusive multiparticle final states (e.g., $\gamma 4\pi$,
$\gamma 4K$, $\gamma K^0K^0$ and $\gamma K^0K\pi$) would help elucidate
the nature of such radiative decays.

\section*{Acknowledgments}
We gratefully acknowledge the effort of the CESR staff
in providing us with excellent luminosity and running conditions.
D.~Cronin-Hennessy and A.~Ryd thank the A.P.~Sloan Foundation.
This work was supported by the National Science Foundation,
the U.S. Department of Energy, and
the Natural Sciences and Engineering Research Council of Canada.

\end{document}